\documentclass[11pt]{article}

\usepackage{array}
\usepackage{subcaption}
\usepackage{subdepth}
\usepackage{url}
\usepackage{fullpage} 
\usepackage{graphicx}
\usepackage{authblk}
\usepackage{amsmath}
\usepackage{amsfonts}
\usepackage{algorithm}
\usepackage{algorithmicx}
\usepackage{algpseudocode}

\title{An Efficient Compression Method for Sign Information of\\ DCT Coefficients via Sign Retrieval}

\author[1]{Chihiro Tsutake}
\author[1]{Keita Takahashi}
\author[1]{Toshiaki Fujii}

\affil[1]{
Department of Information and Communication Engineering, 
Nagoya University, 
Furo-cho, 
Chikusa-ku, 
Nagoya, 
464-8603, 
Japan}

\date{\empty}

\begin{document}

\maketitle
\vspace{-10mm}

\begin{abstract} 
Compression of the sign information of discrete cosine transform coefficients is an intractable problem in image compression schemes due to the equiprobable occurrence of the sign bits. To overcome this difficulty, we propose an efficient compression method for such sign information based on phase retrieval, which is a classical signal restoration problem attempting to find the phase information of discrete Fourier transform coefficients from their magnitudes. In our compression strategy, the sign bits of all the AC components in the cosine domain are excluded from a bitstream at the encoder and are complemented at the decoder by solving a sign recovery problem, which we call \textit{sign retrieval}. The experimental results demonstrate that the proposed method outperforms previous techniques for sign compression in terms of a rate-distortion criterion. Our method implemented in Python language is available from \url{https://github.com/ctsutake/sr}.
\end{abstract}

\noindent
{\bf Keywords.} Image coding, discrete cosine transform, sign information, phase retrieval, sign retrieval, compressive sampling, convex optimization.

\section{Introduction}
The discrete cosine transformation (DCT)~\cite{ANR74} is known as an essential building block for image compression technology that can approximately represent a natural image by using a reduced number of coefficients in the cosine domain~\cite{CP84}. Actual image compression methods, e.g., JPEG, first divide an original image into blocks and then apply the DCT to the blocks followed by a quantization process. Entropy coding is finally performed for efficient binary representations of the quantization indices, in which we face a compression problem for the sign information due to the equiprobable occurrence of the sign bits~\cite{LG00}. We address this sign compression problem for DCT coefficients in this paper.

We briefly summarize seminal works developed to tackle this challenging problem. Tu and Tran~\cite{TT02} proposed a technique inspired by sub-band coding in which DCT coefficients of specific frequency bands are collected in new blocks, and the sign bits are then compressed efficiently by exploiting their correlation in a new block, as was done in EBCOT~\cite{Taubman00}. Ponomarenko et al.~\cite{PBE07} proposed a two-stage technique, where the pixel values of a target block are first estimated from those of already encoded and decoded neighbor blocks and then used to predict the sign bits of DCT coefficients. Since there is a possibility that several predicted sign bits will be incorrect, they proposed transmitting the residual information, i.e., the prediction error, comprising compressible binaries. Clare et al.~\cite{CHJ11} proposed a compression technique known as sign data hiding in which the encoding of one sign bit for each $4 \times 4$ sub-block can be avoided under a mild condition. There are many other works in addition to those mentioned above, e.g., \cite{Memon98,LLS98,Lakhani13}.

To achieve more efficient compression of the sign bits, we propose a novel method based on an approach significantly different from the current literature. Our method relies on a classical signal processing method, referred to as phase retrieval (PR)~\cite{KST95} detailed in Section~\ref{s2}. Specifically, we first retrieve the sign information of all the AC components in the cosine domain only from their magnitudes, and the encoder and decoder then respectively compose and parse a tailored bitstream that includes the residual information to compensate for errors in the retrieved signs, as was done in \cite{PBE07}. To the best of our knowledge, this is the first paper that incorporates the PR methodology into image compression. We show through experiments that our method outperforms the previous ones in terms of a rate-distortion criterion.

\subsection{Contributions}
\label{s1ss2}
The main contributions of this paper are threefold:
\begin{itemize}
    \item A mathematical formulation of a non-convex retrieval problem for the sign information in DCT-based image compression, which we call \textit{sign retrieval} (SR).\vspace{-1mm}
    \item A derivation of a convex relaxation of the SR problem based on theories of recently developed phase retrieval~\cite{GS18} and compressive sampling (CS)~\cite{CENR11}.\vspace{-1mm}
    \item Construction of an efficient convex solution method that combines a classical Fienup algorithm~\cite{Fienup82} with a cascading technique.
\end{itemize}
Note that Salehi et al.~\cite{SBH18} proposed a CS-based convex phase retrieval technique similar to our method. They used a set of Gaussian measurements~\cite{TAB18} to guarantee a theoretical performance, but block-wise cosine measurements were not considered. Our study is thus located on a different position from the measurement viewpoint.

\subsection{Paper Organization}
\label{s1ss3}
The rest of this paper is organized as follows. In Section~\ref{s2}, we summarize a classical PR problem~\cite{KST95} and its solution~\cite{GS18}. Section~\ref{s3} elaborates the proposed method. In Section~\ref{s4}, we report the experimental results and discuss the advantages of our method. We conclude in Section~\ref{s5} with a brief summary and mention of future work.

\subsection{Notations}
\label{s1ss4}
Throughout this paper, we denote a 2D image and its pixels by $\mathbf{x}$ and $x_{n_1,n_2}$, respectively, where $n_1$ and $n_2$ are horizontal and vertical indices, respectively. Let $\mathbf{x}_{b_1,b_2}$ and $x_{b_1,b_2;i_1,i_2}$ be a block of the image $\mathbf{x}$ and its pixels, respectively, where $b_1$ and $b_2$ are block indices, and $i_1$ and $i_2$ are indices in the block. Let $\hat{x}_{k_1,k_2}$ and $\tilde{x}_{b_1,b_2;u_1,u_2}$ be the DFT and DCT coefficients of the image $\mathbf{x}$ and the block $\mathbf{x}_{b_1,b_2}$, respectively; $(k_1,k_2)$ and $(u_1,u_2)$ are indices in the Fourier and cosine domains, respectively. We assume that image and block sizes are $N_1 \times N_2$ and $B_1 \times B_2$, respectively. The inner product operator and the $\ell_p$ norm are denoted by $\langle\cdot,\cdot\rangle$ and $\|\,\cdot\,\|_p$, respectively.
\section{Phase Retrieval}
\label{s2}
\begin{figure}[!t]
\centering
\begin{minipage}[!t]{0.49\linewidth}
\centering
\includegraphics[width=70mm]{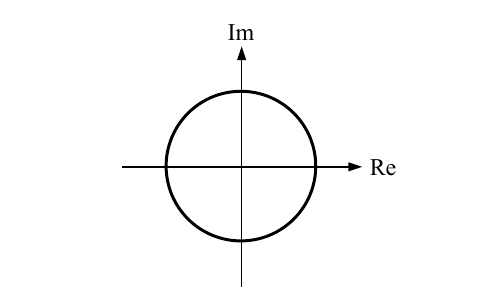}
\subcaption{Non-convex circle in \eqref{eq:pr}}
\end{minipage}
\begin{minipage}[!t]{0.49\linewidth}
\centering
\includegraphics[width=70mm]{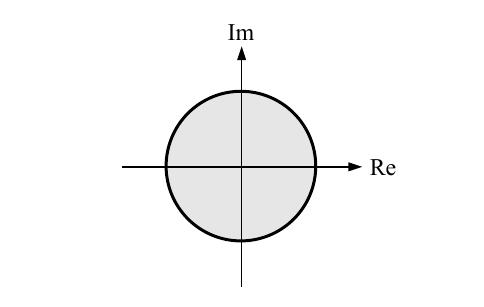}
\subcaption{Convex disk in \eqref{eq:pm}}
\end{minipage}
\caption{Constraints in the Fourier domain.}
\label{fig:const_p}
\end{figure}

Phase retrieval~\cite{KST95} is a classical signal processing method that attempts to solve a non-convex problem of the form 
\begin{equation}
    \mathop{\mathrm{find}} \quad \mathrm{phase}(\hat{x}_{k_1,k_2}) \quad
    \mathrm{from} \quad |\hat{x}_{k_1,k_2}|,\quad \forall_{k_1,k_2}.
    \label{eq:pr}
\end{equation}
Note that \eqref{eq:pr} cannot be solved within polynomial time because of its NP-hardness~\cite{FMNW14}. To avoid this difficulty, Goldstein and Studer~\cite{GS18} proposed a convex relaxation of \eqref{eq:pr}, referred to as PhaseMax:
\begin{equation}
    \mathbf{z}^*=\mathop{\mathrm{argmax}}_{\mathbf{z} \in \mathbb{R}^{N_1 \times N_2}}\,\,\,
    \langle \mathbf{z}, \boldsymbol{\phi}\rangle \quad \mathrm{s.t.} \quad
    |\hat{z}_{k_1,k_2}| \leq |\hat{x}_{k_1,k_2}|,\quad \forall_{k_1,k_2},
    \label{eq:pm}
\end{equation}
where $\boldsymbol{\phi} \in \mathbb{R}^{N_1 \times N_2}$ is the so-called anchor vector. Figure~\ref{fig:const_p} shows a geometrical interpretation of constraints in the Fourier domain, where (a) and (b) correspond to \eqref{eq:pr} and \eqref{eq:pm}, respectively. We have to find the phase information from a non-convex circle in \eqref{eq:pr}, while \eqref{eq:pm} allows the retrieval in a convex disk.

Goldstein and Studer showed the surprising result that \eqref{eq:pr} and \eqref{eq:pm} are equivalent if the number of samples $|\hat{x}_{k_1,k_2}|$ is higher than the number of pixels, i.e., oversampled, and $\boldsymbol{\phi}$ is chosen in such a way that the constant
\begin{equation}
    \alpha(\mathbf{x}, \boldsymbol{\phi}) = 1 - \frac{2}{\pi} \cdot 
    \arccos\bigg( 
    \frac{\langle \mathbf{x}, 
    \boldsymbol{\phi}\rangle}{\|\mathbf{x}\|_2 \cdot \|\boldsymbol{\phi}\|_2} \bigg)
\end{equation}
is sufficiently large. In other words, the original image $\mathbf{x}$ including the exact phase information can be recovered from a set of magnitude measurements in the Fourier domain.

\section{Proposed method}
\label{s3}
\subsection{Encoder and Decoder}
\label{s3ss1}
To achieve efficient compression of the sign information, we modify a bitstream standardized in a conventional DCT-based image compression method, e.g., JPEG. We construct the following encoder and decoder, in which $\tilde{y}_{b_1,b_2;u_1,u_2}$ denotes a degraded version of the original DCT coefficient $\tilde{x}_{b_1,b_2;u_1,u_2}$ obtained by means of the quantization process in the conventional image compression method.\vspace{2mm}\\
\noindent
\textbf{Encoder}: The sign information of all the AC components in the cosine domain are excluded from a bitstream. In other words, only the magnitudes $|\tilde{y}_{b_1,b_2;u_1,u_2}|$ are first compressed by suitable entropy coding. The encoder then performs SR defined in the next subsection to retrieve the sign information. Finally, since there is a possibility that several retrieved signs will be incorrect, we append the residual information
\begin{equation}
    e_{b_1,b_2;u_1,u_2}=
    \mathrm{sgn}(\tilde{y}_{b_1,b_2;u_1,u_2}) \oplus \mathrm{ret\_\,sgn}_{b_1,b_2;u_1,u_2}
    \label{eq:res}
\end{equation}
to the bitstream, where $\mathrm{ret\_\,sgn}_{b_1,b_2;u_1,u_2}$ is the retrieved sign and $\oplus$ is the XOR operator.\vspace{2mm}\\
\noindent
\textbf{Decoder}: The decoder first parses the bitstream to obtain the magnitudes $|\tilde{y}_{b_1,b_2;u_1,u_2}|$ and the residues $e_{b_1,b_2;u_1,u_2}$. The sign information is then retrieved via SR using only $|\tilde{y}_{b_1,b_2;u_1,u_2}|$ in the same way as in the encoder. Finally, the retrieved signs including errors are corrected by $e_{b_1,b_2;u_1,u_2}$.\vspace{2mm}\\
Note that the residual information hopefully has many zeros but few ones, and is thus compressible, if the sign bits are retrieved correctly to some extent. In fact, we can recover almost all the sign bits correctly, as will be shown in Section~\ref{s4}.
\subsection{Sign Retrieval and Its Convex Relaxation}
\label{s3ss2}
\begin{figure}[!t]
\centering
\begin{minipage}[!t]{0.49\linewidth}
\centering
\includegraphics[width=70mm]{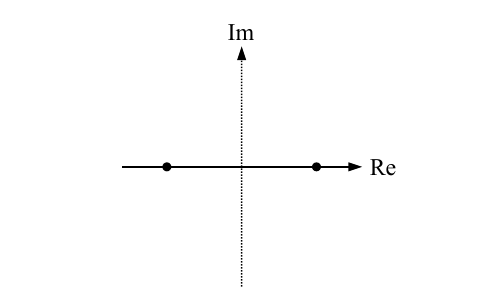}
\subcaption{Non-convex points in \eqref{eq:sr}}
\end{minipage}
\begin{minipage}[!t]{0.49\linewidth}
\centering
\includegraphics[width=70mm]{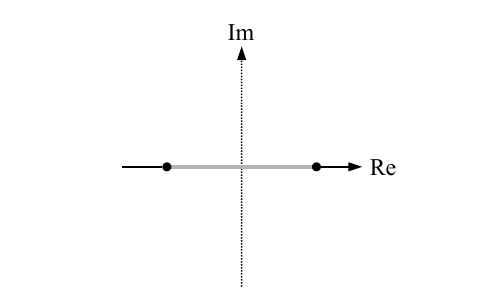}
\subcaption{Convex line in \eqref{eq:sm}}
\end{minipage}
\caption{Constraints in the cosine domain.}
\label{fig:const_s}
\end{figure}

According to \eqref{eq:pr}, retrieval of the sign information can be formulated as a non-convex optimization problem:
\begin{equation}
    \mathrm{find} \quad
    \mathrm{sgn}(\tilde{y}_{b_1,b_2;u_1,u_2})\quad
    \mathrm{from} \quad |\tilde{y}_{b_1,b_2;u_1,u_2}|,\quad
    \forall_{b_1,b_2,u_1,u_2},
    \label{eq:sr}
\end{equation}
which we call \textit{sign retrieval} in our image compression scenario. As in \cite{IK07}, it can be interpreted as a special case of \eqref{eq:pr} by replacing the phase information and the Fourier magnitude with the sign information and the cosine magnitude, respectively. Motivated by this, we formulate a convex relaxation of \eqref{eq:sr} based on \eqref{eq:pm} as follows:
\begin{equation}
    \mathbf{z}_{b_1,b_2}^*=
    \mathop{\mathrm{argmax}}_{\mathbf{z}_{b_1,b_2} \in \mathbb{R}^{B_1 \times B_2}}\,\,\,
    \langle \mathbf{z}_{b_1,b_2}, \boldsymbol{\phi}_{b_1,b_2} \rangle\quad
    \mathrm{s.t.}\quad |\tilde{z}_{b_1,b_2;u_1,u_2}| \leq 
    |\tilde{y}_{b_1,b_2;u_1,u_2}|,\quad\forall_{b_1,b_2,u_1,u_2}.
    \label{eq:sm}
\end{equation}
Figure~\ref{fig:const_s} shows constraints of \eqref{eq:sr} and \eqref{eq:sm}, similarly to Fig.~\ref{fig:const_p}. There are no imaginary parts in the cosine domain, so that the constraint in \eqref{eq:sm} is represented as a convex real line.

We then apply the CS theory~\cite{CENR11} to \eqref{eq:sm} because of its ill-posedness from the sampling viewpoint: the DFT coefficients are oversampled in \eqref{eq:pm} while the number of DCT coefficients is equivalent to that of the pixels in \eqref{eq:sm}. The theory claims that a sparse signal can be restored from a set of few measurements by solving a constrained $\ell_1$-norm minimization problem. In our compression setting, $|\tilde{y}_{b_1,b_2;u_1,u_2}|$ can be seen as the measurements in the cosine domain, so that we introduce an $\ell_1$ norm function in \eqref{eq:sm}. Let $\boldsymbol{\Psi}$ be a semi-orthonormal matrix satisfying $\boldsymbol{\Psi}^\mathrm{t}\boldsymbol{\Psi}=\mathrm{Id}$ for promoting the sparsity of the entire image $\mathbf{z}$. We formulate a regularized SR problem as
\begin{equation}
    \mathbf{z}^* = 
    \mathop{\mathrm{argmax}}_{\mathbf{z} \in \mathbb{R}^{N_1 \times N_2}}\quad
    \sum_{b_1,b_2} \langle \mathbf{z}_{b_1,b_2}, \boldsymbol{\phi}_{b_1,b_2}\rangle - 
    \lambda \|\boldsymbol{\Psi} \mathbf{z}\|_1
    \quad\mathrm{s.t.}\quad
    |\tilde{z}_{b_1,b_2;u_1,u_2}| \leq 
    |\tilde{y}_{b_1,b_2;u_1,u_2}|,\quad\forall_{b_1,b_2,u_1,u_2},
    \label{eq:rsm}
\end{equation}
where $\lambda$ is a regularization parameter. By solving this, note that the retrieved sign can be given as
\begin{equation*}
    \mathrm{ret\_\,sgn}_{b_1,b_2;u_1,u_2} =
    \mathrm{sgn}(\tilde{z}_{b_1,b_2;u_1,u_2}^*),
\end{equation*}
where $\tilde{z}_{b_1,b_2;u_1,u_2}^*$ is a DCT coefficient of a block in $\mathbf{z}^*$.
\subsection{Solution to \eqref{eq:rsm}}
\label{s3ss3}
We elaborate hereafter an actual solution method for \eqref{eq:rsm}. We apply the Fienup algorithm~\cite{Fienup82}, a conventional but efficient solution method for the PR problem~\eqref{eq:pr}, to our SR. The original Fienup algorithm features two building blocks: proximal operations~\cite{CP10} and projection onto non-convex constraint sets (Fig.~\ref{fig:const_p}(a)).

With this in mind, since the constraint is convex in our problem, we first introduce the corresponding convex set 
\begin{equation*}
    \mathcal{C}_{b_1,b_2;u_1,u_2} = 
    \big\{\tilde{z}_{b_1,b_2;u_1,u_2}:|\tilde{z}_{b_1,b_2;u_1,u_2}| \leq |\tilde{y}_{b_1,b_2;u_1,u_2}|\big\}.
\end{equation*}
Let $\mathcal{C}$ denote a union of all the set $\mathcal{C}_{b_1,b_2;u_1,u_2}$, and let
\begin{eqnarray*}
    \hspace{3mm}
    \imath_{\mathcal{C}}\big(z\big) = 
    \begin{cases}
       \hspace{0.7mm}0 \quad \mathrm{if} \quad 
       \tilde{z}_{b_1,b_2;u_1,u_2} \in \mathcal{C}, \,\,\,{}^\forall b_1,b_2,u_1,u_2\\
       \infty \hspace{2.5mm} \mathrm{o/w}
    \end{cases}\hspace{-3mm}
\end{eqnarray*}
be an indicator function. The Fienup algorithm for \eqref{eq:rsm} can be described as the following three steps by performing variable splitting~\cite{ABF10}:
\begin{align}
    \mathbf{f}_{[\theta+1]}^* &= 
    \mathop{\mathrm{argmin}}_{\mathbf{f} \in \mathbb{R}^{N_1 \times N_2}}\,
    \lambda\|\boldsymbol{\Psi} \mathbf{f}\|_1 + 
    \frac{1}{2}\Big\|\mathbf{f} - \mathbf{z}_{[\theta]}^*\Big\|_2^2
    \label{eq:sub1}\\
    \mathbf{g}_{[\theta+1]}^* &= \mathop{\mathrm{argmin}}_{\mathbf{g} \in \mathbb{R}^{N_1 \times N_2}}\,
    -\langle \mathbf{g}, \boldsymbol{\phi} \rangle + 
    \frac{\mu}{2}\Big\|\mathbf{g} - \mathbf{f}_{[\theta+1]}^*\Big\|_2^2
    \label{eq:sub2}\\
    \mathbf{z}_{[\theta+1]}^* &= \mathop{\mathrm{argmin}}_{\mathbf{z} \in \mathbb{R}^{N_1 \times N_2}}\,
    \imath_{\mathcal{C}}(\mathbf{z}) 
    + \frac{1}{2}\Big\|\mathbf{z} - \mathbf{g}_{[\theta+1]}^*\Big\|_2^2,
    \label{eq:sub3}
\end{align}
where $\theta$ is the number of iterations, and $\mu$ is a positive penalty parameter. Note that \eqref{eq:sub1}, \eqref{eq:sub2}, and \eqref{eq:sub3} are proximal operations in terms of $\|\,\cdot\,\|_1$, $\langle\cdot,\cdot\rangle$, and $\imath_\mathcal{C}$, respectively. According to~\cite{CP10}, we have the closed-form solutions of these problems as follows:
\begin{align}
    \mathbf{f}_{[\theta+1]}^* &= 
    \boldsymbol{\Psi}^\mathrm{t}
    \Big\{ \mathrm{sgn} \Big(\boldsymbol{\Psi}\mathbf{z}_{[\theta]}^*\Big) 
    \cdot \Big(
    \big|\boldsymbol{\Psi} 
    \mathbf{z}_{[\theta]}^*\big|-\lambda\Big)_+\Big\}
    \label{eq:sol1}\\
    \mathbf{g}_{[\theta+1]}^* &= \mathbf{f}_{[\theta+1]}^* + \frac{1}{\mu} \boldsymbol{\phi}
    \label{eq:sol2}\\
    \mathbf{z}_{[\theta+1]}^* &= \mathrm{proj}_{\mathcal{C}}\Big(\mathbf{g}_{[\theta+1]}^*\Big),
    \label{eq:sol3}
\end{align}
where $(a)_+ = \max(a, 0)$, and $\mathrm{proj}_{\mathcal{C}}$ is the projection operator onto $\mathcal{C}$. The iteration is performed until $\theta$ reaches the maximum count $\Theta$.

Finally, we discuss how to determine the anchor vector $\boldsymbol{\phi}$. In PR, because $\boldsymbol{\phi}$ is chosen so that $\alpha(\mathbf{x},\boldsymbol{\phi})$ is sufficiently large, we also determine $\alpha(\mathbf{y}, \boldsymbol{\phi})$ on the basis of this criterion. To this end, we first use a random initialization method, as in \cite{GS18}, and then solve \eqref{eq:rsm} repeatedly by setting $\boldsymbol{\phi} = \mathbf{z}^*$, where $\mathbf{z}^*$ is the solution of the former iteration. This is because $\mathbf{z}^*$ would be a better approximation of $\mathbf{y}$, leading to a higher $\alpha(\mathbf{y}, \boldsymbol{\phi})$. We cascade the solution $\Gamma$ times, and the entire algorithm is described as Algorithm~\ref{alg:cfa}.

\begin{figure}[!t]
\vspace{-3mm}
\begin{algorithm}[H]
\renewcommand{\algorithmicrequire}{\textbf{Input:}}
\renewcommand{\algorithmicensure}{\textbf{Output:}}
\caption{Cascaded Fienup algorithm}
\label{alg:cfa}
\begin{algorithmic}
\Require Parameters $\lambda > 0$ and $\mu > 0$ 
\Ensure Image $\mathbf{z}^*$
\State $\boldsymbol{\phi} \leftarrow \mathrm{rand}$
\For{$\gamma=1,\cdots,\Gamma$}
\For{$\theta=1,\cdots,\Theta$}
\State Update $\mathbf{f}_{[\theta+1]}^*$, 
$\mathbf{g}_{[\theta+1]}^*$, 
and $\mathbf{z}_{[\theta+1]}^*$ 
by \eqref{eq:sol1}--\eqref{eq:sol3}.\vspace{0.5mm}
\EndFor
\State $\boldsymbol{\phi} \leftarrow \mathbf{z}^*$
\EndFor
\State \Return $\mathbf{z}^*$
\end{algorithmic}
\end{algorithm}
\end{figure}

\begin{figure}[!t]
\centering
\begin{minipage}[!t]{0.32\linewidth}
\centering
\includegraphics[width=35mm]{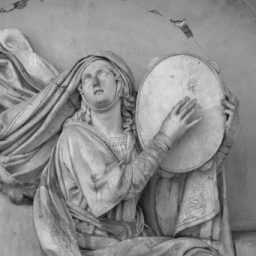}
\subcaption{\textit{Statue}}
\end{minipage}
\begin{minipage}[!t]{0.32\linewidth}
\centering
\includegraphics[width=35mm]{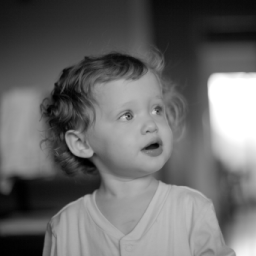}
\subcaption{\textit{Baby}}
\end{minipage}
\begin{minipage}[!t]{0.32\linewidth}
\centering
\includegraphics[width=35mm]{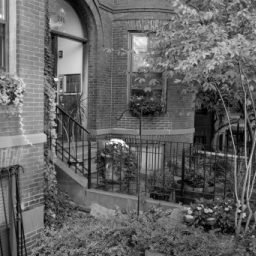}
\subcaption{\textit{House}}
\end{minipage}
\caption{Original images.}
\label{fig:org}
\end{figure}

\begin{figure}[!t]
\centering
\begin{minipage}[!t]{0.99\linewidth}
\centering
\hspace{-13mm}
\includegraphics[width=82mm]{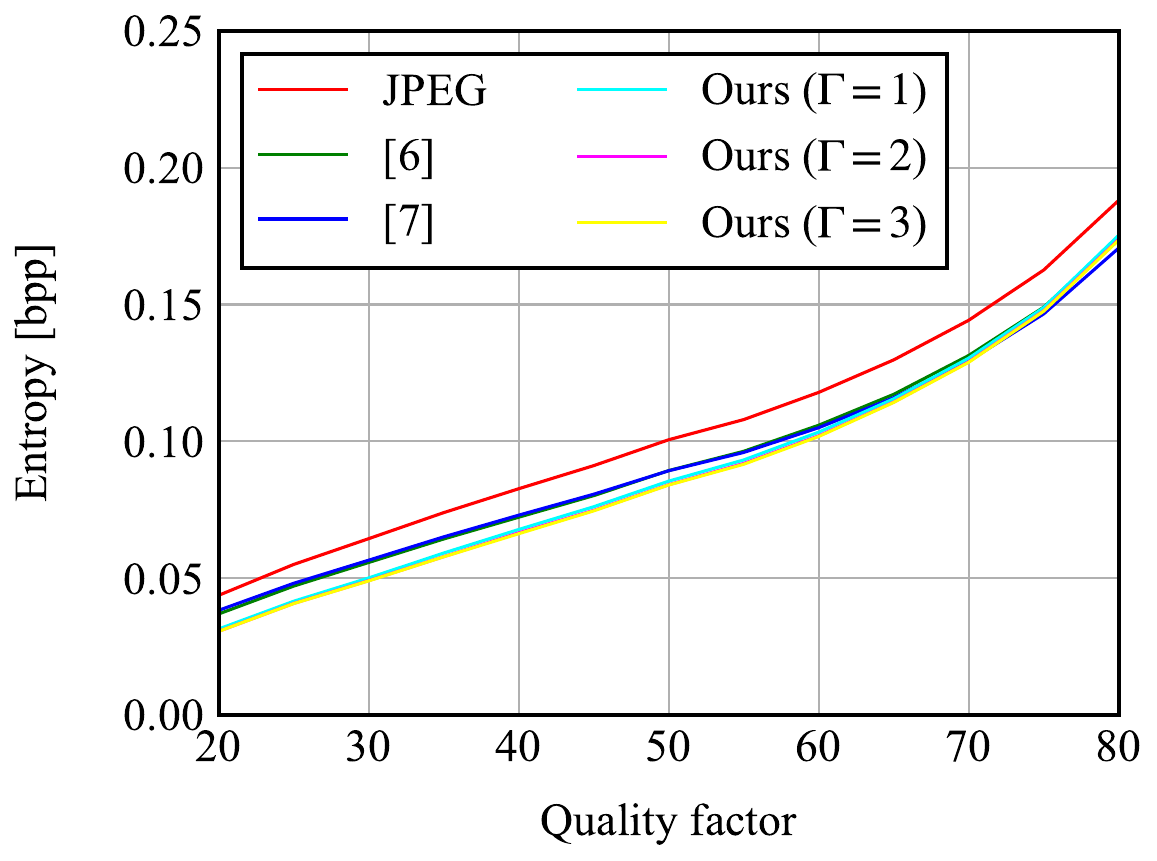}
\subcaption{\textit{Statue}}
\end{minipage}
\begin{minipage}[!t]{0.99\linewidth}
\centering
\hspace{-13mm}
\includegraphics[width=82mm]{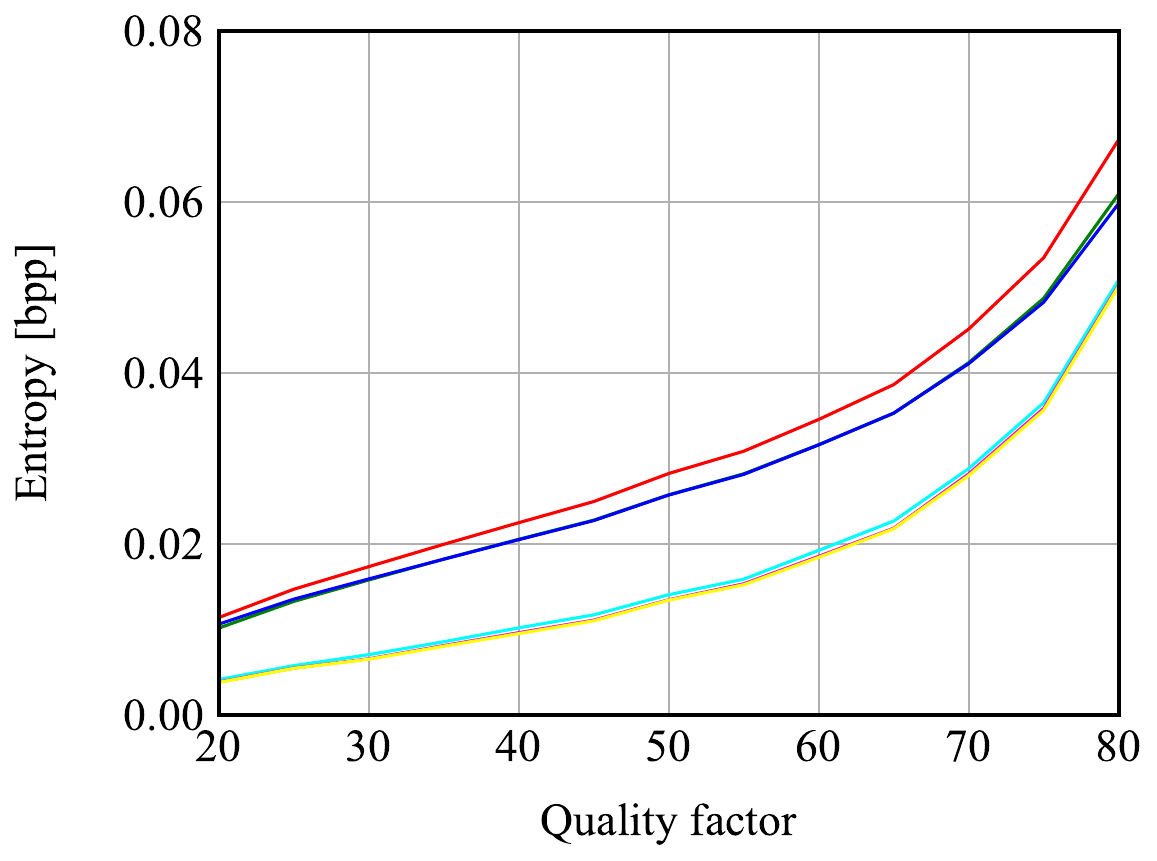}
\subcaption{\textit{Baby}}
\end{minipage}
\begin{minipage}[!t]{0.99\linewidth}
\centering
\hspace{-13mm}
\includegraphics[width=82mm]{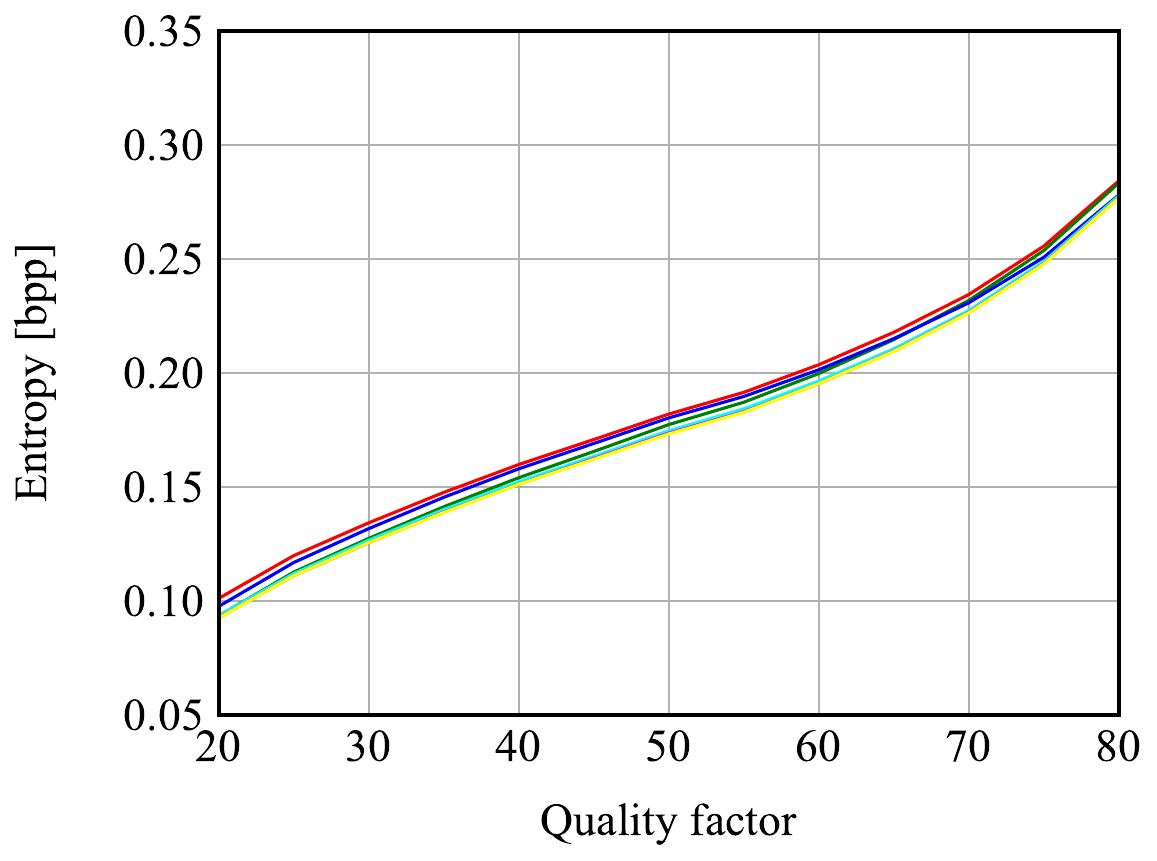}
\subcaption{\textit{House}}
\end{minipage}
\caption{Entropy for sign information of DCT coefficients.}
\label{fig:res}
\end{figure}

\begin{figure}[!t]
\centering
\begin{minipage}[!t]{0.32\linewidth}
\centering
\includegraphics[width=35mm]{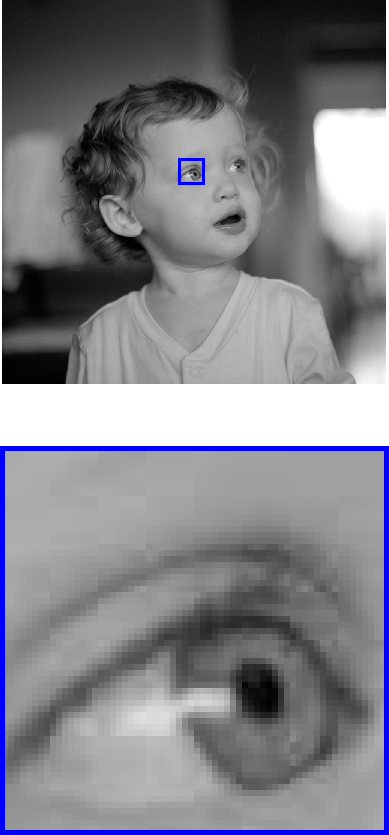}
\subcaption{Baseline}
\end{minipage}
\begin{minipage}[!t]{0.32\linewidth}
\centering
\includegraphics[width=35mm]{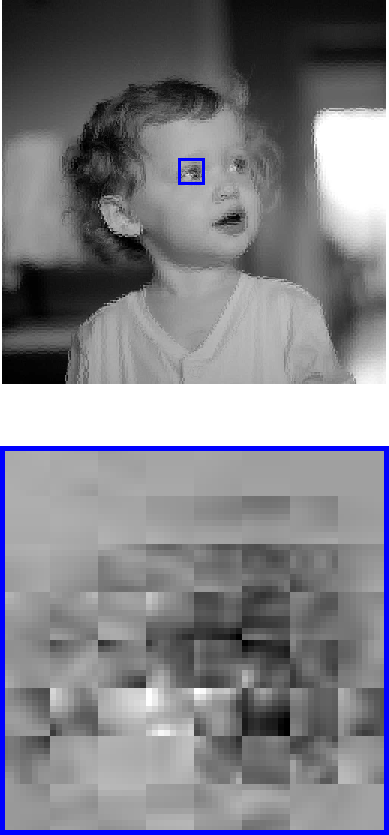}
\subcaption{Random signs}
\end{minipage}
\begin{minipage}[!t]{0.32\linewidth}
\centering
\includegraphics[width=35mm]{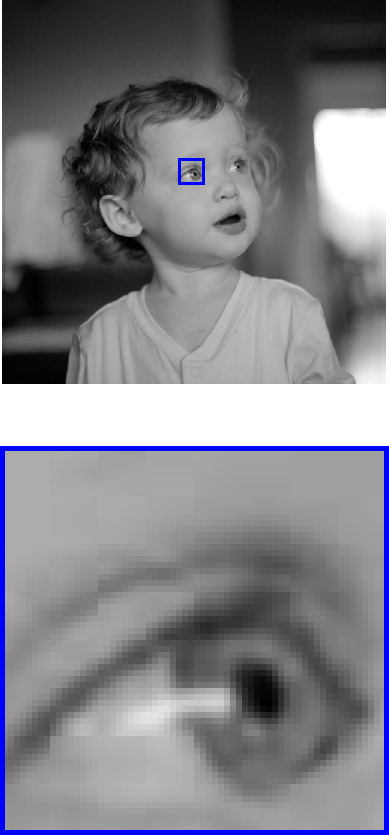}
\subcaption{Ours ($\Gamma=3$)}
\end{minipage}
\caption{Examples of reconstructed images.}
\label{fig:rec}
\end{figure}
\section{Experimental Results}
\label{s4}
To evaluate the effectiveness of our method, we implemented it in JPEG with $B_1 \times B_2=8 \times 8$ and compared it with \cite{PBE07} and \cite{CHJ11}. We used the original images shown in Fig.~\ref{fig:org}, which were generated by cropping $N_1 \times N_2 = 1024 \times 1024$ regions of images from the MIT-Adobe FiveK Dataset~\cite{BPCD11}. All the images were compressed with the quality factor varied from $20$ to $80$. Because all the methods yield the same error against the original image, we evaluated the entropy [bpp] required for compressing the sign bits, e.g., $e_{b_1,b_2;u_1,u_2}$ in our method. 

The parameters $\lambda$ and $\mu$ were set to $1$ and $0.01$, respectively. We ran $\Theta=200$ iterations and varied the number of cascading $\Gamma$ from $1$ to $3$ because it significantly affects the accuracy of $\mathrm{ret\_\,sgn}_{b_1,b_2;u_1,u_2}$. A translation invariant version~\cite{CD95} of the $12$-th order Symmlet transformation~\cite{Daubechies92} was used as $\boldsymbol{\Psi}$, leading to $\boldsymbol{\Psi}^\mathrm{t}\boldsymbol{\Psi}=\mathrm{Id}$.

Figure~\ref{fig:res} shows the entropy required to transmit the sign information of the DCT coefficients by each method. Compared to the previous techniques, our method with $\Gamma=1$ had a lower entropy. This result can be improved by increasing $\Gamma$, which demonstrates that cascading is effective in our method. Figure~\ref{fig:rec} shows several examples of reconstructed images, where (a) is a JPEG image and (b) and (c) are reconstructed images using random sign bits and those retrieved by Algorithm~\ref{alg:cfa} with $\Gamma= 3$, respectively. While (b) was completely degraded, (c) was very close to (a), indicating that most of the sign bits were correctly retrieved. Note that in our method, the errors in (b) and (c) were eventually corrected by using $e_{b_1,b_2; u_1, u_2}$ to obtain exactly the same image as (a). However, when used with Algorithm~\ref{alg:cfa}, the residual bits consisted of many zeros and few ones, resulting in small entropy values for the sign information, as shown in Fig.~\ref{fig:res}. These results demonstrate the advantages of the proposed method over the previous techniques from the rate-distortion point of view.
\section{Conclusion}
\label{s5}
In this paper, we addressed an intractable sign compression problem for DCT coefficients in image compression by developing a method based on PR. Specifically, we first formulated a PR-based non-convex optimization problem, i.e., SR, and its convex relaxation by means of the $\ell_1$-norm regularization. The resulting optimization problem was solved by the cascaded Fienup algorithm. Our method was validated by a comparison with previous techniques. We presume our method can be applied to any DCT-based compression standard, such as HEVC~\cite{SOHW12} or VVC~\cite{BCL19}, which we will verify in future work.

\bibliographystyle{IEEEtran}
\bibliography{main}

\end{document}